\documentclass[english,aps,prb,superscriptaddress,reprint,longbibliograhy]{revtex4-2}

\usepackage{graphicx}  
\usepackage{float}     
\usepackage{dcolumn}   
\usepackage{bm}        
\usepackage{amssymb}   
\usepackage{amsmath}
\usepackage{lipsum}    
\usepackage{xcolor}
\usepackage{soul}
\usepackage{float}
\usepackage{placeins}
\usepackage{physics}
\usepackage{hyperref}
\hypersetup{
    colorlinks=true,     
    linkcolor=blue,      
    citecolor=blue,      
    urlcolor=blue,       
    }

\usepackage[utf8]{inputenc}
\usepackage{natbib}
\usepackage{graphicx}
\usepackage{subcaption}
\usepackage{booktabs}
\usepackage{blindtext}
\usepackage{color}

\definecolor{goodred}{RGB}{183,15,58}
\definecolor{goodblue}{RGB}{93,128,180}

\begin{document}

\title{\Large\textbf{Theory of tensorial magnetic inertia in terahertz spin dynamics} }

\author{Subhadip Ghosh}
\email[]{22dr0226@iitism.ac.in}
\affiliation{Department of Physics, Indian Institute of Technology (ISM) Dhanbad, IN-826004, Dhanbad, India}

\author{Mikhail Cherkasskii}
\affiliation{Institute for Theoretical Solid State Physics, RWTH Aachen University, 52074 Aachen, Germany}

\author{Igor Barsukov}
\affiliation{Department of Physics and Astronomy, University of California, Riverside, California 92521, USA}
\author{Ritwik Mondal}
\email[]{ritwik@iitism.ac.in}
\affiliation{Department of Physics, Indian Institute of Technology (ISM) Dhanbad, IN-826004, Dhanbad, India}

\begin{abstract}
     Magnetic inertia has emerged as a possible way to manipulate ferromagnetic spins at a higher frequency e.g., THz. Theoretical treatments so far have considered the magnetic inertia as a scalar quantity. Here, we explore the magnetic inertial dynamics with a magnetic inertia tensor as macroscopic derivations predicted it to be a tensor. First, the inertia tensor has been decomposed into three terms: (a) scalar and isotropic inertia, (b) anisotropic and symmetric inertia tensor, (c) chiral and antisymmetric tensor. Further, we employ linear response theory to the inertial Landau-Lifshitz-Gilbert equation with the inertia tensor and calculate the effect of chiral and anisotropic inertia on ferromagnets, antiferromagnets, and ferrimagnets. It is established that the precession and nutation resonance frequencies decrease with scalar magnetic inertia. Our results suggest that the nutation resonance frequencies further reduce due to inertia tensor. However, the effective damping of the nutation resonance increases with the chiral and antisymmetric part of the inertia tensor.  We show that the precession resonances remain unaffected, while the nutation resonances are modified with the chiral magnetic inertia. 
    
\end{abstract}
\maketitle

\section{Introduction}
Understanding femtosecond and picosecond manipulations of magnetic spins is crucial for developing faster and more efficient magnetic storage devices and other spintronic applications \cite{Beaurepaire1996,Kimel2007-Review,Kirilyuk2010}. A magnetic material having a resonance frequency in the THz regime becomes an extremely prominent candidate e.g., antiferromagnet (AFM) \cite{Nemec2018,Jungwirth2016}. The AFM materials are characterized by having no net magnetization because their magnetic moments are aligned in opposite directions, canceling each other out. This results into smaller domain sizes compared to ferromagnetic materials \cite{Fitzsimmons2008,krizek2022Science}. They pose challenges for certain applications, such as magnetic storage, where larger and more distinct domains are often preferable for data stability and readability. 

An alternative way of ultrafast spin manipulation that can be achieved in a ferromagnet (FM) at THz frequencies has been demonstrated recently using the spin inertia \cite{neeraj2019experimental,Wegrowe2015JAP}. The presence of spin inertia breaks the parallel alignment of magnetic moment and angular momentum \cite{Kikuchi,Wegrowe2012,Mondal2020nutation,cherkasskii2020nutation,Li2022APL}. Consequently, the magnetic moment rotates around the angular momentum. Such rich spin dynamics can be theoretically modelled by the  inertial Landau-Lifshitz-Gilbert (ILLG) equation \cite{Ciornei2011,Titov2021Deterministic,Mondal2017Nutation,Cherkasskii2022Anisotropy,MONDAL_Review}
\begin{align}
        \frac{\partial \bm{M}}{\partial t} = & \bm{M} \times \left[-\gamma \bm{H}_{\rm eff} +   \frac{\alpha}{M_0} \frac{\partial \bm{M}}{\partial t} + \frac{\eta}{M_0} \frac{\partial^2 \bm{M}}{\partial t^2}\right]\,.
        \label{Eq:0}
\end{align}
Here, the first term in the right side explains the magnetization precession around an effective field  $\bm{H}_{\rm eff}$. Without the last term in Eq. (\ref{Eq:0}), the frequency of the magnetization precession becomes $\gamma \bm{H}_{\rm eff}$, with $\gamma$ as the gyromagnetic ratio. The second term represents the isotropic Gilbert damping with dimensionless damping parameter $\alpha$, while the third term delineates the magnetic inertial dynamics with a magnetic inertial relaxation time $\eta$  \cite{Olive2012,Wegrowe2015JAP,Fahnle2011,fahnle2013erratum}. Notably, the inertial spin dynamics have been experimentally observed in NiFe, CoFeB, and Co films \cite{neeraj2019experimental,unikandanunni2021inertial,de2024nutationseparatingspinmagnetic}. 

The microscopic origin of Gilbert damping has been attributed to several mechanisms including relativistic spin-orbit coupling \cite{hickey09,Mondal2016,Fahnle2011JPCM}, breathing Fermi surface model and torque-torque correlation model \cite{kambersky70,kambersky07,Thonig_2014,Thonig2018,gilmore07}, scattering theory and linear response theory \cite{tserkovnyak02,EbertPRL2011}. Importantly, these works define that the Gilbert damping parameter is a tensor \cite{Brinker_2022,Guimaraes_2019,Nagyfalusi2024,Osorio2024}. 

Similarly, the microscopic origin of magnetic inertia has been explored using the torque-torque correlation model \cite{Thonig2017,Bajaj2024}, incorporating the extended breathing Fermi surface model \cite{Fahnle2011,fahnle2013erratum}, along with the Keldysh formalism \cite{Bhattacharjee2012}, a time-dependent non-equilibrium Green's function approach \cite{Bajpai2019}, and relativistic Dirac-Kohn-Sham formalism \cite{Mondal2017Nutation,Mondal2018JPCM}. The findings in Ref. \cite{Fahnle2011,Mondal2017Nutation,Bajaj2024,Juba2019PRM} demonstrated the tensorial nature of magnetic inertia. However, this tensorial inertia is often simplified to a scalar form, which limits a deeper understanding.

While the experimental observations of the inertial dynamics are limited to FM \cite{neeraj2019experimental,unikandanunni2021inertial,de2024nutationseparatingspinmagnetic}, the signatures of magnetic inertia has been predicted theoretically using a scalar and isotropic magnetic inertia characterised by $\eta$ in collinear FM \cite{Cherkasskii2021,Cherkasskii2022Anisotropy,titov2022ferromagnetic,Titov2022NutationWaves,MondalJPCM2021,Rodriguez2024PRL,Adams2024PRB}, AFM \cite{Mondal2020nutation,Mondal2021PRB,Titov2024JAP,He2023,He2024PRB} and non-collinear conical spin spirals \cite{CherkasskiiPRB2024}. The magnetic inertial dynamics introduces a second resonance peak in the frequency response spectrum in a FM. Such resonance is termed as nutation resonance \cite{cherkasskii2020nutation}. While the precession resonance occurs at GHz frequencies, the nutation resonance occurs at THz frequencies, thereby providing an opportunity to manipulate FM spins at faster timescales e.g. picoseconds \cite{Giordano2020,Lomonosov2021,Makhfudz2020,Makhfudz2022}. It has been shown that the nutation resonance is exchange enhanced in AFM compared to the same in FM \cite{Mondal2020nutation,Mondal2021PRBSpinCurrent}. Due to the magnetic inertial dynamics, the emergence of nutational wave has been predicted that has lower group velocity compared to the precessional spin wave \cite{Titov2022NutationWaves,Lomonosov2021,Mondal2022,Makhfudz2020}. It has also been reported that nutational 90$^{\rm o}$ switching can be achieved by using a linearly polarised THz pulse, while 180$^{\rm o}$ switching can be achieved by using a circularly polarised THz pulse     \cite{Winter2022,Makhfudz2022}. The inertial switching time is reduced in AFM, compared to FM \cite{Kimel2009,Winter2022}. 

All the above theoretical and computational studies have used magnetic inertia as a scalar quantity. However, several investigations have anticipated it to be a tensor \cite{Fahnle2011,Mondal2017Nutation,Bajaj2024,Thonig2017}. Here, we explore the effect of magnetic inertial dynamics with the magnetic inertia parameter as a tensor $\Delta$ in FM, AFM, and ferrimagnets (FiM). First, we decompose the inertia tensor $\Delta$ into three parts: (a) isotropic and scalar characterized by $\eta$, (b) anisotropic and symmetric tensor characterized by $\mathbb{I}$, and (c) chiral and antisymmetric tensor characterized by a vector $\bm{C}$. We, then, use a linear response theory to compute the magnetic susceptibility as a function of frequency. By finding the poles of the susceptibility, we obtain the precession and nutation resonance frequencies, which are generally complex. It is already known that both these frequencies decrease with scalar magnetic inertia $\eta$. Along the same line, the nutation resonance frequencies further decrease due to inertia tensor, while the precession resonance frequency remains almost unaffected. An interesting feature of the chiral magnetic inertia is that the effective damping of the nutation resonance increases sharply, providing an opportunity to manipulate the nutation resonance. In contrast, the effective damping of the precession resonance is decreased with chiral inertia. Our direct solutions match quite well with the approximated expression obtained with linear order terms for FM, however, the higher-order terms in the analytical expressions cannot be ignored for AFM. Following the trends in AFM, we also find that the precession resonance is unaffected with inertia tensor in FiM, However, the nutation resonance frequency is decreased in FiM.   

The rest of the paper is organized as follows: Sec. \ref{Section2} introduces the decomposition of inertia tensor and linear response theory applied to one and two sublattice magnetic systems. We discuss the effect of inertia tensor in FM, AFM and FiM in the Sec. \ref{Section3A}, \ref{Section3B} and \ref{Section3C}, respectively. Especially, we focus on the chiral magnetic inertia in these systems. In Sec. \ref{Section4}, we conclude with a summary of our work.    

\section{Methods}\label{Section2}
    The generalized ILLG equation with the Gilbert damping tensor and inertia tensor can be written as  \cite{Fahnle2011,Mondal2017Nutation,Bajaj2024,Nagyfalusi2024}
\begin{align}\label{eq:1}
        \frac{\partial \bm{M}}{\partial t} = &-\gamma \bm{M} \times \bm{H}_{\rm eff} +   \frac{\bm{M}}{M_{0}}\times \left({\Gamma} \cdot \frac{\partial \bm{M}}{\partial t}\right)\nonumber\\ 
        & + \frac{\bm{M}}{M_{0}}\times \left({\Delta}\cdot \frac{\partial^2 \bm{M}}{\partial t^2}\right)\,.
\end{align}
 
The terms in the ILLG spin dynamics in Eq. (\ref{eq:1}) comprise of spin precession around an effective field $\bm{H}^{\rm eff}$ ({ in the unit of Tesla}) where $\gamma/2\pi = 28$ GHz/T is the gyromagnetic ratio, the Gilbert damping with a damping tensor ${\Gamma}$, and the magnetic inertial dynamics with the inertia tensor ${\Delta}$. Note that the Gilbert damping is a dimensionless quantity, however, the magnetic inertia has a dimension of time. These tensors can be decomposed into symmetric and antisymmetric parts. This decomposition has already been worked out in Ref. \cite{Dhali_2024} for Gilbert damping tensor $\Gamma$. We follow a similar decomposition for the inertia tensor $\Delta$. The symmetric and antisymmetric parts of the tensor $\Delta$ can be written as $\Delta^{\rm sym}_{ij} = \frac{1}{2}\left(\Delta_{ij} + \Delta_{ji}\right)$ and $ \Delta^{\rm antisym}_{ij} = \frac{1}{2}\left(\Delta_{ij} - \Delta_{ji}\right)$, respectively. The antisymmetric part $\Delta^{\rm antisym}_{ij}$ can further be expressed in terms of a vector $\bm{{ C}}$ as $\Delta^{\rm antisym}_{ij} = \epsilon_{ijk}{C}_k$, with $\epsilon_{ijk}$ as Levi-Civita symbol \cite{Mondal2018PRB}. On the other hand, the symmetric part $\Delta^{\rm sym}_{ij}$ can further be decomposed as $\Delta^{\rm sym}_{ij} = \eta\delta_{ij} + \mathbb{I}_{ij}$ with $\delta_{ij}$ as  Kroneker-delta and $\mathbb{I}_{ij}$ as a symmetric tensor. With such decomposition of the magnetic inertia tensor, the ILLG spin dynamics in Eq. (\ref{eq:1}) can be recast as
\begin{align}
& \frac{\partial {\bm M}}{\partial t} = -\gamma\bm{M} \times {\bm H}_{\rm eff} + \frac{\alpha}{M_{0}}\left({\bm M} \times \frac{\partial {\bm M}}{\partial t}\right) \nonumber\\
& + \frac{\bm{M}}{M_0}\times \left[\eta \frac{\partial^2 \bm{M}}{\partial t^2} +\left(\mathbb{I}\cdot \frac{\partial^2 \bm{M}}{\partial t^2}\right) -\left(\bm{{ C}}\times\frac{\partial^2 \bm{M}}{\partial t^2}\right) \right]\, . 
\label{eq:2}
\end{align}
Note that only the scalar part $\alpha$ has been considered for the Gilbert damping tensor. The contributions of the other parts of the tensorial Gilbert damping have been discussed extensively in Ref. \cite{Dhali_2024}. The decomposition of the inertia tensor $\Delta$ leads to three terms in the ILLG spin dynamics {\it viz.} the second line of Eq. (\ref{eq:2}). The first term explains scalar and isotropic inertial dynamics with inertial relaxation time $\eta$ which is a scalar quantity. The second term depicts the anisotropic inertial dynamics with a symmetric inertia tensor $\mathbb{I}$. The last term signifies the chiral inertial dynamics with an inertia vector $\bm{C}$.

In this work, we focus on the effect of anisotropic and symmetric inertial dynamics and the chiral and antisymmetric inertial dynamics for ferromagnets and antiferromagnets. We exploit the linear response theory to calculate the eigen modes, resonance frequencies, and effective damping in these model magnetic systems. 

For one sublattice FM, we consider that the ground state is characterised by $\bm{M} = M_0 \hat{\bm{z}}$ and it is under the influence of an external field $\bm{H} = H_0\hat{\bm{z}}$. The free energy of the system can written as $\mathcal{F}(\bm{M}) = -H_0M_z - KM_z^2/M_0^2$ with $K$ as the uniaxial anisotropy energy. The effective field entering in ILLG spin dynamics of Eq. (\ref{eq:2}) can be calculated as $\bm{H}_{\rm eff} = -\partial \mathcal{F}/\partial \bm{M} = \left(H_0 + 2KM_z/M_0^2\right)\hat{\bm{z}}$. A small in-plane oscillating field $\bm{h}(t)$ generates the induced magnetization $\bm{m}(t)$ such that the total magnetization becomes time dependent $\bm{M}(t) = m_x(t)\hat{\bm{x}} + m_y(t)\hat{\bm{y}} + M_0\hat{\bm{z}}$ where $m_x, m_y << M_0$. We switch to the circularly polarized basis, $m_{\pm} = m_x \pm \textrm{i}m_y = me^{\pm i\omega t}$ and $h_{\pm} = h_x \pm \textrm{i}h_y =he^{\pm i\omega t}$ and keeping only the linear-order response terms in the ILLG spin dynamics, we explore the contributions of magnetic inertia tensor in FM.

AFMs and FiM have two sublattices that align antiparallel to each other. We assume that the magnetizations of the two sublattices $A$ and $B$ are $\bm{M}_A = M_{A0} \hat{\bm{z}}$ and $\bm{M}_B = - M_{B0} \hat{\bm{z}}$ at the ground state. Considering the homogeneous magnetization for AFM, the free energy is written as 
\begin{align}
       & \mathcal{F}(\bm{M}_A,\bm{M}_B) = \nonumber\\ & -H_0(M_{Az}+M_{Bz}) - \frac{K_A}{M^2_{A0}}M^2_{Az}  - \frac{K_B}{M^2_{B0}}M^2_{Bz} \nonumber\\
       & + \frac{1}{M_{A0}M_{B0}} \left[ J\bm{M}_A\cdot\bm{M}_B +\bm{D}\cdot \left(\bm{M}_A\times \bm{M}_B\right)\right],
\label{eq:3}
\end{align}
which include the external field $\bm{H} = H_0\Hat{\bm{e_z}}$, uniaxial anisotropy energies $K_A$, $K_B$ and the Heisenberg exchange interaction energy $J$. We further include the last energy term in Eq. (\ref{eq:3}) due to antisymmetric exchange -- so called Dzyaloshinskii-Moriya interaction (DMI) through the DM vector $\bm{D}$ \cite{DZYALOSHINSKY1958,Moriya1960}. For AFM  and FiM, one has to solve two ILLG equations for two sublattices $A$ and $B$, including the inertia tensor. The effective fields that enter into the ILLG equations can be obtained as $\bm{H}_{A/B} = -{\partial\mathcal{F}(\bm{M}_A,\bm{M}_B)}/{\partial\bm{M}_{A/B}}$. These fields are calculated as 
\begin{align}
    \bm{H}_{A/B} = &\left(H_0\,+\frac{2K_{A/B}M_{A/Bz}}{M^2_{A/B0}}\right)\hat{\bm{z}}\nonumber\\
    &-\frac{1}{{M_{A0}M_{B0}}}\left[{J \,\bm{M}_{A/B} + \bm{M}_{A/B}\times \bm{D}} \right]\, .
\label{eq:4}
\end{align}
Similar to the ferromagnetic case, we assume small in-plane time-dependent fields $\bm{h}_A(t)$ and  $\bm{h}_B(t)$ induce magnetizations in AFM such that $\bm{M}_A(t) =  {m}_{Ax}(t)\bm{\hat{x}} +{m}_{Ay}(t)\bm{\hat{y}} +M_{A0}\bm{\hat{z}} $ and  $\bm{M}_B(t) ={m}_{Bx}(t)\bm{\hat{x}} + {m}_{By}(t)\bm{\hat{y}} -M_{B0}\bm{\hat{z}}$. We switch to the circularly polarized basis spanned by $m_{A/B\pm} = m_{A/Bx} \pm {\rm i}m_{A/By} \propto e^{\pm {\rm i}\omega t}$ and $h_{A/B\pm} = h_{A/Bx} \pm \textrm{i}h_{A/By}\propto e^{\pm {\rm i}\omega t}$ and compute the effect of inertia tensor in AFM. The corresponding results are presented in the next section.   

Note that a similar linear response theory has been applied to ILLG spin dynamics with scalar inertia $\eta$ in Ref. \cite{Mondal2020nutation}. Therefore, we explore the anisotropic inertial dynamics via the symmetric tensor $\mathbb{I}$ and the chiral inertial dynamics via the antisymmetric part represented by a vector $\bm{C}$. While  we consider $\bm{C} = C\hat{\bm{z}}$ for a FM,  $\bm{C}_A = C\hat{\bm{z}}$ and $\bm{C}_B = -C\hat{\bm{z}}$ have been assumed for AFM and FiM. The magnitude of the off-diagonal elements of the symmetric tensor $\mathbb{I}_{xy}$ is assumed smaller than that of the diagonal elements $\mathbb{I}_{xx}$ and $\mathbb{I}_{yy}$. Moreover, the effects of $\mathbb{I}_{xy}$ are also smaller.      
   
\section{Results and Discussions}
\subsection{Ferromagnets}\label{Section3A}
Using the linear response theory discussed earlier, the susceptibility for a FM can be calculated as
\begin{align}\label{eq:5}
     \chi_{\pm}& = \frac{\gamma M_0}{\Omega_0 - \omega -\left(\eta +  \mathbb{I}_{xx}\right)\omega^2 \pm {\rm i}\left(\alpha + C \omega\right)\omega}
\end{align}
with $\Omega_0 = \gamma\left(2K + H_0 M_0\right)/M_0$. Note that we have considered $\mathbb{I}_{xy} = \mathbb{I}_{yx} = 0$.  The dissipated power can be computed through $P_{\rm FM} ={\dot{\bm{m}}} \cdot \bm{h} = \omega {\tt Im}(\chi_\pm)\vert h\vert^2$, where ${\tt Im}$ represents the imaginary part.
\begin{figure}[tbh!]
    \centering
\includegraphics[scale = 0.3]{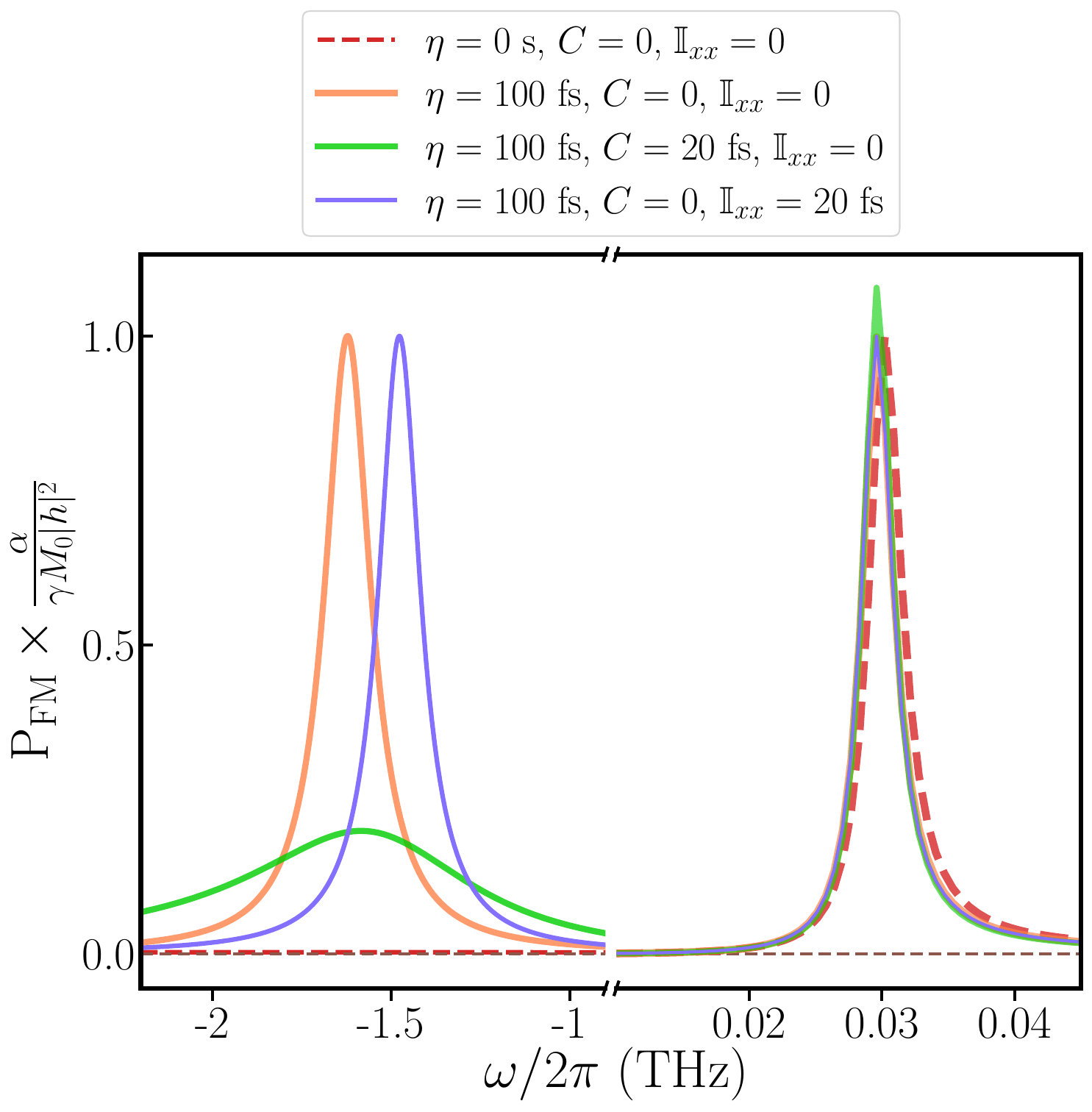}
    \caption{Dissipated power has been computed as a function of frequency at different values of inertial relaxation time. The considered parameters are $\gamma/2\pi$ = 28 GHz/T, $M_0 = 2\mu_B$, $\alpha = 0.05$, $K = 10^{-23}$ J, $H_0 = 0$ T. 
    }
    \label{fig:1}
\end{figure}
The calculated dissipated power has been shown in Fig. \ref{fig:1}. In non-inertial case, only the ferromagnetic resonance (FMR) in the GHz regime is obtained \cite{Farle1998,Kittel1948}. In our case, we obtained the FMR frequency at 30 GHz approximately with the considered model parameters. It is known that the scalar inertial relaxation time $\eta$ introduces an additional resonance peak at the THz frequency regime - this is called nutation resonance (see orange curve in Fig. \ref{fig:1}) \cite{Wegrowe2015JAP,cherkasskii2020nutation,Mondal2020nutation,Mondal2021PRB}. With the value of $\eta$ = 100 fs, the additional nutation peak is obtained about 1.7 THz. When we fixed $\eta = 100$ fs and applied ${\mathbb I}_{xx} = 20$ fs, the FMR frequency remains same. But, the nutation resonance frequency is red-shifted. The reason being the nutation resonance frequency scales with $1/(\eta + \mathbb{I}_{xx})$. A salient feature of the tensorial inertia can be observed with the chiral inertial dynamics governed by the vector $\bm{{C}}$. At $\eta = 100$ fs and $C = 20$ fs we find a rather wider nutation resonance peak where the effective damping of nutation resonance is enhanced. We further analyze the anisotropic and chiral inertial dynamics in detail for FMs.

To obtain the resonance frequencies, one has to find the poles of the susceptibility leading to
\begin{align}
     \left(\eta +  \mathbb{I}_{xx} \mp {\rm i}C\right) \omega^2 +  (1 \mp {\rm i}\alpha) \omega -\Omega_0 = 0\,.
     \label{eq:6}
\end{align}
Therefore, 
The real parts (denoted by ${\tt Re}(\omega)$) of the precession ($\omega_{\rm p}$) and nutation ($\omega_{\rm n}$) frequencies are approximated as
\begin{align}
\label{Eq7}
\omega_{\rm p} & \approx  \Omega_0\left(1 - \beta_{\rm FM}\right),  \\
\omega_{\rm n}  & \approx - \frac{\eta^{\rm eff}}{\left(\eta^{\rm eff}\right)^2 + C^2} + \Omega_0 \left(1-\beta_{\rm FM}\right),
\label{Eq8}
\end{align}
with $\eta^{\rm eff} = \left(\eta + \mathbb{I}_{xx}\right)$ and $\beta_{\rm FM} = \eta^{\rm eff} \Omega_0$. The approximation is obtained using the linear-order terms in the inertia tensor. Note that when considering only the scalar inertia $\eta$, the leading term in nutation frequency $\omega_{\rm n}$ is $-1/\eta$ as investigated in Ref. \cite{cherkasskii2020nutation}.   

\subsubsection{Effect of anisotropic inertial dynamics}
To understand the effect of anisotropic inertial dynamics, we set $C = 0$ and directly solve Eq. (\ref{eq:6}).
\begin{figure}[tbh!]
    \centering
\includegraphics[scale = 0.4]{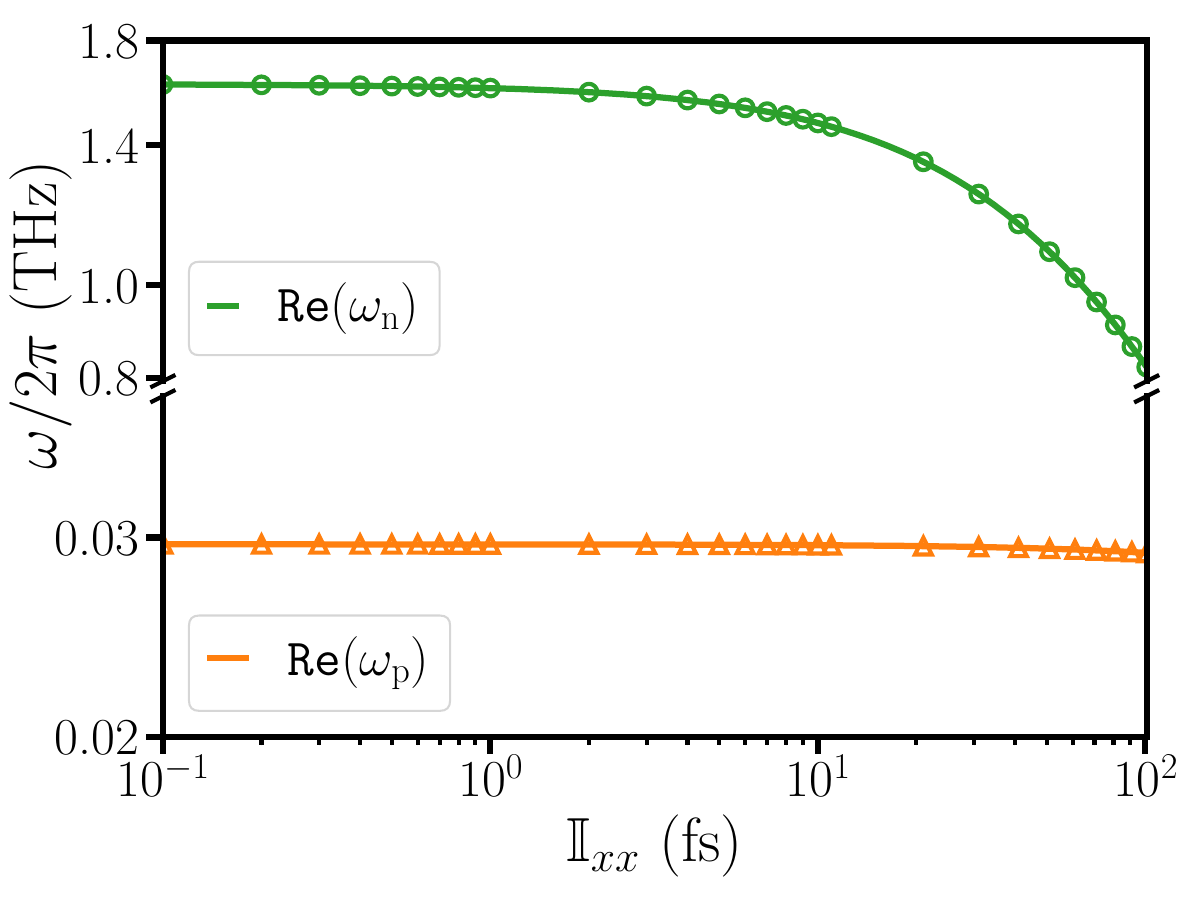}
    \caption{The variation of precession and nutation frequencies plotted against $\mathbb{I}_{xx}$. The considered parameters are $\gamma/2\pi$ = 28 GHz/T, $M_0 = 2\mu_B$, $\alpha = 0.05$, $K = 10^{-23}$ J, $H_0 = 0$ T. 
}
\label{fig:2}
\end{figure}
Hereafter, the inertial relaxation time $\eta$ has been adapted as $\eta = 100$ fs throughout this work, unless specified. Such a value is consistent with the experimental value of $\eta$ reported about 300 fs in CoFeB \cite{neeraj2019experimental}, while 120 fs and 110 fs reported in epitaxial fcc and bcc Co films, respectively \cite{unikandanunni2021inertial}. We restrict the values of the anisotropic inertial relaxation time $\mathbb{I}_{xx}$ such that $\mathbb{I}_{xx}<\eta$. The computed precession and nutation frequencies are shown in  Fig. \ref{fig:2}. 
\begin{figure}[h]
    \centering
\includegraphics[scale=0.4]{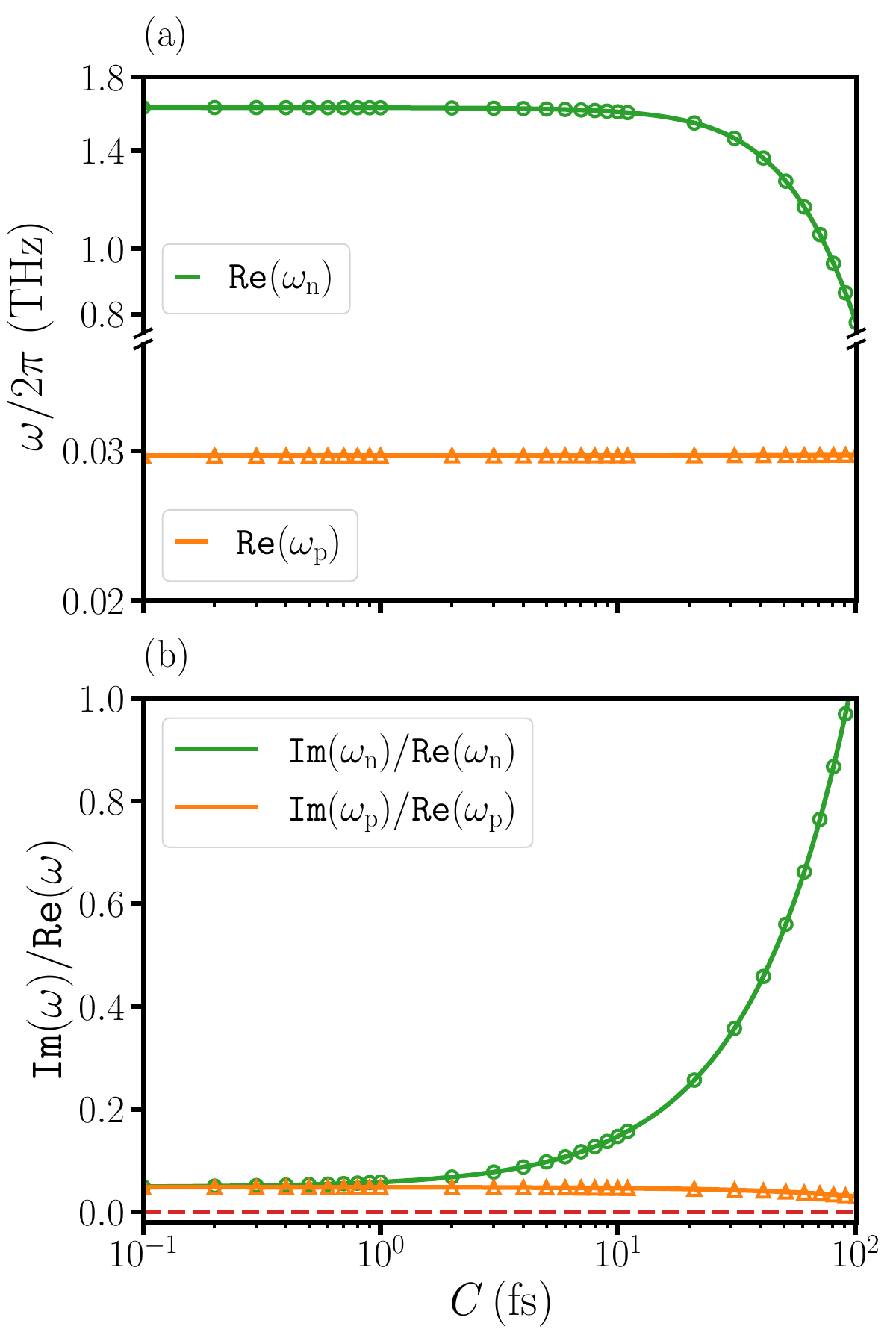}
    \caption{The variation of (a) precession and nutation frequencies and (b) effective damping plotted against $C$. The considered parameters are $\gamma/2\pi$ = 28 GHz/T, $M_0 = 2\mu_B$, $\alpha = 0.05$, $K = 10^{-23}$ J, $H_0 = 0$ T. 
    }
    \label{fig:3}
\end{figure} 
 
The results suggest that the nutation resonance frequency decreases with $\mathbb{I}_{xx}$ quite significantly, however, the precession resonance frequency remains almost constant. This can be attributed to Eqs. (\ref{Eq7}) and (\ref{Eq8}) derived earlier. It is clear that the dominant term in $\omega_{\rm p}$ is $\Omega_0$ $\approx 2\pi \times$ 30 GHz with our model parameters. For the estimation, the leading correction term $\beta_{\rm FM} \Omega_0 \approx 0.04 \Omega_0$ is negligibly small compared to $\Omega_0$. Therefore, the precession resonance frequency barely depends on $\mathbb{I}_{xx}$. On the other hand, the dominant term in $\omega_{\rm n}$ is $1/\eta^{\rm eff}$ with $C = 0$. For smaller $\mathbb{I}_{xx}$, the nutation frequency changes slightly up to 1.83\%. However, higher $\mathbb{I}_{xx}$ significantly decrease the nutation frequency following $1/\eta^{\rm eff}$. The precessional and nutational effective damping, defined as the ratio of the real and imaginary part of the respective frequencies are not affected by $\mathbb{I}_{xx}$.

To summarize, the symmetric part of the tensorial inertia does not influence the precession resonance frequency. However, the nutation resonance frequency is decreased. 

\subsubsection{Effects of chiral inertial dynamics}
Similarly, we use $\mathbb{I}_{xx} = \mathbb{I}_{yy} = 0$ for further analysis of the chiral inertial dynamics characterized by $C\neq 0$. The computed precession and nutation resonance frequencies are found with Eq. (\ref{eq:6}). The solutions are shown in Fig. \ref{fig:3}(a). Evidently, the precession resonance frequency is not influenced by the chiral inertial time $C$. Such results can also be verified by the approximation in Eq.\ (\ref{Eq7}) as it does not depend on $C$ at the leading order. However, the nutation resonance frequency is decreased as $C$ increases. Setting $\mathbb{I}_{xx} = 0$ in Eq. (\ref{Eq8}), the dominant nutation frequency term becomes $-\eta / (\eta^2 + C^2)$. It is quite clear that the nutation frequency decreases as $C$ increases. 

The effective damping ${\tt Im}(\omega)/{\tt Re}(\omega)$ for precession and nutation resonances are shown in Fig. (\ref{fig:3})(b). While the chiral inertial dynamics $C$ decrease the effective damping of the precession resonance, the effective damping of the nutation resonance is enhanced quite significantly. This observation can also be verified from the green curve in Fig. (\ref{fig:1}) where the effective damping can be calculated from the full width at half maxima of each resonance peak. In fact, the effective damping approaches nearly 1 at the limit $ C = \eta$. As the damping signifies the dissipation of energy from spin degrees of freedom to other, the enhancement of damping means the energy dissipation at the nutation will be much more at higher C.

\subsection{Antiferromagnets}\label{Section3B}
Now, we scrutinize the AFMs with two sublattices, namely A and B, and those are antiparallel to each other. 
Following the linear response theory, the susceptibility for AFM becomes a $2\times 2$ matrix
\begin{widetext}
    \begin{align}\label{eq:9}
    \chi^{AB}_{\pm} & = \frac{1}{\Xi_\pm}\begin{pmatrix}
     \dfrac{\Omega_B \pm \textrm{i}\omega\alpha_{B} -\left(\eta_{B} + \mathbb{I}^B_{xx} \mp {\rm i} C_B\right)\omega^2 +\omega }{ \gamma_B M_{B0}} &  -\dfrac{J\mp {\rm i}D_z}{M_{A0}M_{B0}}\\
     - \dfrac{J\pm {\rm i}D_z}{M_{A0}M_{B0}} &    \dfrac{\Omega_A \pm \textrm{i}\omega\alpha_{A} -\left(\eta_{A} + \mathbb{I}^A_{xx} \pm {\rm i} C_A\right)\omega^2-\omega}{ \gamma_A M_{A0}}
    \end{pmatrix}\,,
\end{align}
where 
\begin{align}\label{eq:10}
\Xi_\pm = & \left(\gamma_A M_{A0} \gamma_B M_{B0}\right)^{-1}\left[\Omega_A \pm {\rm i}\omega\alpha_{A} -\left(\eta_{A} + \mathbb{I}^A_{xx} \pm {\rm i} C_A\right)\omega^2-\omega  \right]\left[\Omega_B \pm {\rm i} \omega\alpha_{B} -\left(\eta_{B} + \mathbb{I}^B_{xx} \mp {\rm i} C_B\right)\omega^2 +\omega  \right] \nonumber\\
& - \frac{J^2+ D_z^2}{M_{A0}^2 M_{B0}^2}\,.
\end{align}
\end{widetext}
The frequency of the spin precession for two sublattices can be defined by 
\begin{align}
\Omega_A & = \frac{\gamma_A}{M_{A0}}\left(J+2K_A+H_0M_{A0}\right)\\
\Omega_B  & = \frac{\gamma_B}{M_{B0}}\left(J+2K_B-H_0M_{B0}\right)\,.
\end{align}
These are similar to $\Omega_0$ defined for FM, however due to opposite alignment of two sublattices the Zeeman field contributes positively to one of the suballtices. Switching off the Zeeman field will give rise to same frequency of the two subalttices. 
Note that we have considered the DMI along the equilibrium direction of the magnetization i.e., $\bm{D} = D_z\hat{\bm{e}}_z$. The expression of susceptibility matrix was derived earlier in Refs. \cite{Kamra2018,Mondal2021PRB}, however, without the inertia tensor and antisymmetric DMI. 
For the AFM  resonance frequencies, we calculate the nodes of the susceptibility tensor in Eq. (\ref{eq:9}), obtaining $\Xi_{\pm} = 0$. Therefore, a fourth-order equation in frequency is

\begin{align}
\mathcal{M}_{\pm}\omega^4 + \mathcal{N}_{\pm}\omega^3 + \mathcal{O}_{\pm}\omega^2 + \mathcal{P}_{\pm}\omega + \mathcal{Q}_{\pm} = 0,
\label{Eq:12}
\end{align}
with the coefficients as follows
\begin{align}
\mathcal{M}_{\pm} & = \left(\eta_{A} + \mathbb{I}^A_{xx} \pm {\rm i} C_A\right)\left(\eta_{B} + \mathbb{I}^B_{xx} \mp {\rm i} C_B\right)\,, \\
\mathcal{N}_{\pm} & = \mp {\rm i} \left[\alpha_A\left(\eta_{B} + \mathbb{I}^B_{xx} \mp {\rm i} C_B\right) + \alpha_B \left(\eta_{A} + \mathbb{I}^A_{xx} \pm {\rm i} C_A\right) \right]\nonumber\\
& - \left[\left(\eta_{A} + \mathbb{I}^A_{xx} \pm {\rm i} C_A\right)- \left(\eta_{B} + \mathbb{I}^B_{xx} \mp {\rm i} C_B\right) \right]\,, \\
\mathcal{O}_{\pm} & =  - 1 \pm {\rm i}\left(\alpha_A - \alpha_B\right)  - \alpha_A\alpha_B\nonumber\\
& - \left[ \Omega_A\left(\eta_{B} + \mathbb{I}^B_{xx} \mp {\rm i} C_B\right) + \Omega_B \left(\eta_{A} + \mathbb{I}^A_{xx} \pm {\rm i} C_A\right) \right]\,,\\
\mathcal{P}_{\pm} & = \Omega_A - \Omega_B \pm{\rm i} \left(\Omega_A \alpha_B + \Omega_B \alpha_A\right)\,, \\
\mathcal{Q}_{\pm} & = \Omega_A\Omega_B- \dfrac{\gamma_A \gamma_B \left(J^2 +D^2_z\right)}{M_{A0}M_{B0}} \,.
\end{align}
The solutions of the above equation determine the resonance frequencies. Essentially, one obtains four solutions, two of which are AFM precession resonance frequencies ($\omega_{\rm p\pm}$) while the other two dictate the AFM nutation resonance frequencies ($\omega_{\rm n\pm}$). As these frequencies are complex, we denote $\omega_{\rm p\pm} = {\tt Re}(\omega_{\rm p\pm})+{\rm i}\,{\tt Im}(\omega_{\rm p\pm})$ and $\omega_{\rm n\pm}={\tt Re}(\omega_{\rm n\pm})+{\rm i}\,{\tt Im}(\omega_{\rm n\pm})$.
Further, for simplicity we consider $\eta_A = \eta_B= \eta,\, \alpha_A = \alpha_B = \alpha, \, \mathbb{I}^A_{xx} = \mathbb{I}^B_{xx} = \mathbb{I}_{xx}, \, C_{A} = C_{B} = C, \, \gamma_A = \gamma_B = \gamma, \, M_{A0}= M_{B0}= M_0$ and $K_A = K_B = K$. The direct solution of Eq. (\ref{Eq:12}) has been plotted in Figs. \ref{fig:4} and \ref{fig:5}.  However, we examine an approximate analytical solution of Eq. (\ref{Eq:12}) and compare it with the { direct} solutions. 

Under the approximation $J\gg K, M_0H_0$ and $\alpha\ll1$, we can write $\Omega_A=\Omega_B=\frac{\gamma}{M_0}\left(J+2K\right)$. The fourth-order Eq. (\ref{Eq:12}) with $\mathcal{M}_+$, $\mathcal{N}_+$, $\mathcal{O}_+$, $\mathcal{P}_+$ and $\mathcal{Q}_+$ becomes an effective second-order equation in $\omega^2$ given by  
\begin{align}\label{18}
        &\left[\left(\eta^{\rm eff}\right)^2 +C^2\right]\omega^4- \left[1+\frac{2\gamma\eta^{\rm eff}}{M_0}\left(J+2K\right)\right]\omega^2\nonumber\\
        &-2{\rm i}\left(\alpha\eta^{\rm eff}+C\right)\omega^3_{(0)}+\left[\frac{2i\gamma\alpha}{M_0}\left(J+2K\right)+2\gamma H_0\right]\omega_{(0)}\nonumber\\& 
        +\frac{\gamma^2}{M^2_0}\left(J+2K\right)^2-\gamma^2 H^2_0-\frac{\gamma^2 (J^2 + D_z^2)}{M^2_0}=0,
\end{align}
where $\omega_{(0)}$ is the solution of the equation for $\alpha=0$ and $H_0=0$. We have defined $\eta^{\rm eff} = \eta + \mathbb{I}_{xx}$. Keeping only the linear-order terms in $\alpha$ and $H_0$, as well as $K/J\ll 1$, the precession and nutation resonance frequencies are calculated as
\begin{align}\label{Eq:19}
    &\omega_{\rm p\pm}  \approx \pm \frac{\gamma}{M_0}\sqrt{\frac{4K\left(J+K\right)-D_z^2}{1+\dfrac{2\gamma\eta^{\rm eff}}{M_0}\left(J+2K\right)}} +\frac{\gamma H_0}{1+\dfrac{2\gamma\eta^{\rm eff}}{M_0}\left(J+2K\right)}\nonumber\\& -\frac{\dfrac{{\rm i}\gamma\alpha}{M_0}\left(J+2K\right) -\dfrac{{\rm i}\gamma^2}{M_0^2}\dfrac{\left(\alpha\eta^{\rm eff}+C\right)\left[4K(J+K) - D_z^2\right]}{1+\dfrac{2\gamma\eta^{\rm eff}}{M_0}\left(J+2K\right)}}{1+\dfrac{2\gamma\eta^{\rm eff}}{M_0}\left(J+2K\right)}\\
&\omega_{\rm n\pm} \approx \pm{\sqrt{\frac{1+\dfrac{2\gamma\eta^{\rm eff}}{M_0}(J+2K)}{\left(\eta^{\rm eff}\right)^2+C^2}}}
        \cross\nonumber\\
        &\left[1-\frac{\dfrac{\gamma^2}{M^2_0}\left[4K(J+K)-D_z^2\right]\left[\left(\eta^{\rm eff}\right)^2+C^2 \right]}{2\left(1+\dfrac{2\gamma\eta^{\rm eff}}{M_0}(J+2K)\right)}\right]\nonumber\\
        &-\frac{\gamma H_0+\dfrac{{\rm i}\alpha\gamma}{M_0}(J+2K)}{1+\dfrac{2\gamma\eta^{\rm eff}}{M_0}(J+2K)}+\dfrac{{\rm i}\left(\alpha\eta^{\rm eff}+C\right)}{\left(\eta^{\rm eff}\right)^2+C^2}\label{Eq:20}\, .
\end{align}
Note that the leading-order frequencies have been used as $\omega_{(0)}$ in the respective calculations. While the approximated solutions in Eqs. (\ref{Eq:19}) and (\ref{Eq:20}) have been obtained by setting $\Xi_+ = 0$, the complex conjugation will formulate the solutions for $\Xi_- = 0$. { The} same expressions for precession and nutation resonance frequencies have been obtained in Ref. \cite{Mondal2020nutation}, however with only scalar spin inertia and without the DMI. Analogous to the dimensionless parameter $\beta_{\rm FM}$ defined for ferromagnets, a similar dimensionless parameter can also be defined for AFM as $\beta_{\rm AFM} =  \frac{\gamma\eta^{\rm eff}}{M_0} (J + 2K)$. In comparison to the FM case, the dimensionless parameter in AFM is exchange enhanced by the exchange energy $J$.  In the preceding sections, we thoroughly compare the results via direct solutions of Eq. (\ref{Eq:12}) and the obtained approximated expressions.   

\subsubsection{Effect of anisotropic inertial dynamics}
\begin{figure}[h]
    \centering    \includegraphics[scale=0.42]{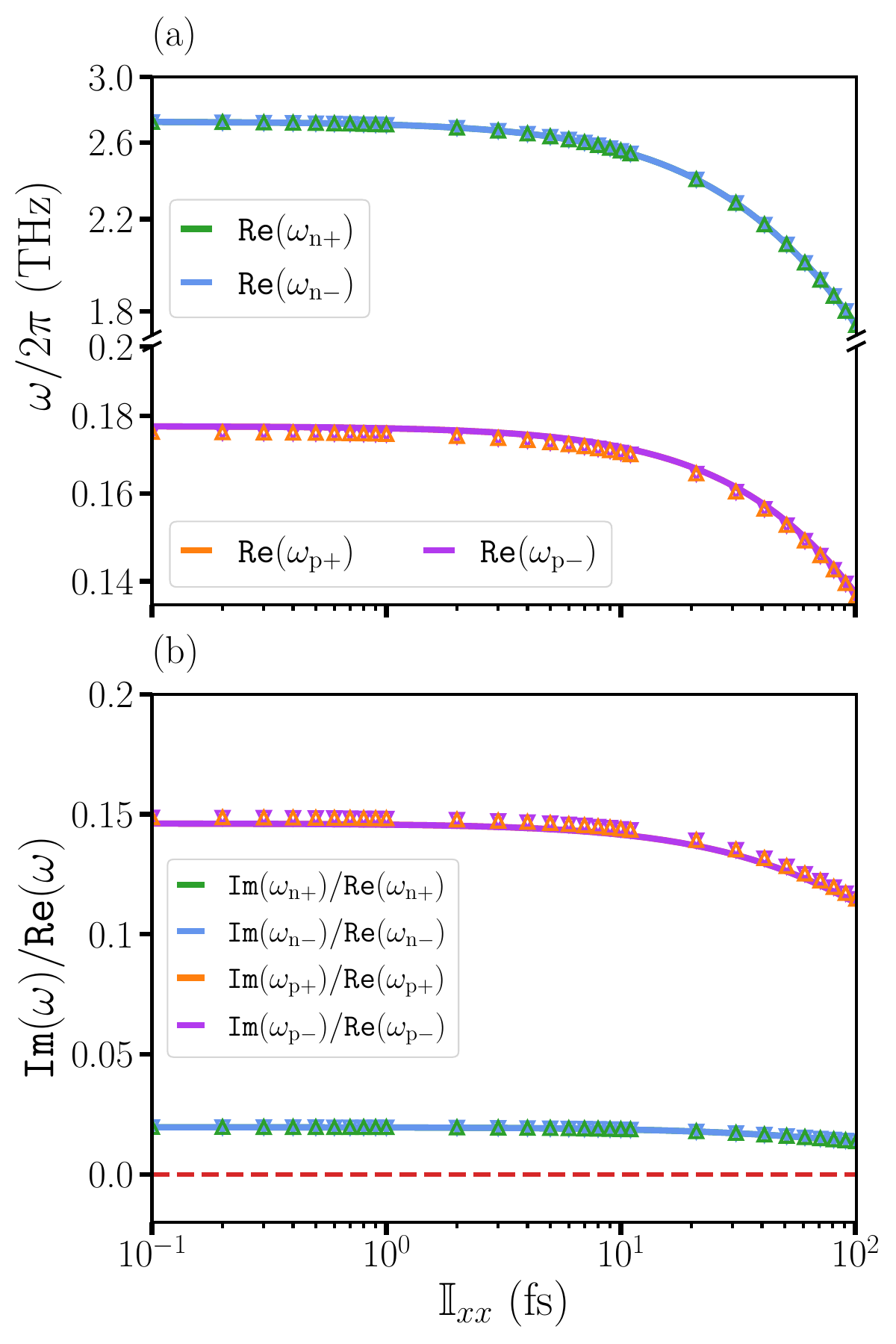}
    \caption{The variation of (a) precession and nutation frequencies and (b) effective damping plotted against $\mathbb{I}_{xx}$. The solid lines represent the approximated expressions derived in Eqs. (\ref{Eq:19}) and (\ref{Eq:20}), whereas the symbols are calculated data points { via the direct solutions of Eq. (\ref{Eq:12})}. The chiral inertia has been set to $C = 0$. The considered parameters are $M_{0}$ = $2\mu_B$, $\gamma$ = 28 GHz/T, $\alpha$= 0.05, $ J = 10^{-21}$ J, $K = 10^{-23}$ J, $D_z = 0$ J, $H_0 = 0$ T. }
    \label{fig:4}
\end{figure}
To understand the effect of $\mathbb{I}_{xx}$, we have set $C = 0$ in the { direct} solutions of Eq. (\ref{Eq:12}) as well as the approximate analytical solutions of Eqs. (\ref{Eq:19}) and (\ref{Eq:20}). We use $\eta = 100$ fs in the solutions presented. Note that the { experimental value of} scalar inertial relaxation time $\eta$ is not known for an AFM, so far. The computed results for resonance frequencies and effective damping are shown as a function of $\mathbb{I}_{xx}$ in Fig. \ref{fig:4}. 

The results suggest that the precession and nutation resonance frequencies decrease with $\mathbb{I}_{xx}$. This trends can be explained by Eq. (\ref{Eq:19}). The dominant real part in precession frequency is the first term in Eq. (\ref{Eq:19}). Similarly, the dominant real part of the nutation frequency is also the first term in Eq. (\ref{Eq:20}). Both these frequencies have $\eta^{\rm eff} = \eta + \mathbb{I}_{xx}$ in the denominator. Therefore, both the resonance frequencies decrease with increasing $\mathbb{I}_{xx}$. Note that both the precession resonance frequencies i.e., $\omega_{\rm p+}$ and $\omega_{\rm p-}$ overlap with each other. This is due to the fact that the applied Zeeman field has been set to zero, $H_0 = 0$. The occurance of two precession resonances in AFM can also be observed even without the magnetic inertia, however at $H_0 \neq 0$ \cite{Kamra2018,Rezende2019,Kittel1951,Titov2024JAP}. We compare the frequencies of resonances in FM and AFM. The precessional resonance $\omega_{\rm p}$ varies by 1.73\% with $\mathbb{I}_{xx}$ in FM, and by 20.52\% in AFM for the parameters listed in Figs. \ref{fig:2} and \ref{fig:4}. The variation of nutation resonance $\omega_{\rm n}$  with $\mathbb{I}_{xx}$  is similar for both FM (49.31\%) and AFM (35.8\%).

 It is known that the effective damping { ${\tt Im}(\omega)/{\tt Re}(\omega)$}  for a precession resonance is exchange enhanced, even without the magnetc inertia \cite{Moriyama2017PRL,Kamimaki2019,Chiba2015PRB,Baltz2018}. The presence of scalar magnetic inertia $\eta$ decreases the effective damping of the precession as well as nutation resonances \cite{Mondal2020nutation}. Here, we find that the effective damping decreases with the tensorial inertia. Our obtained results show very good agreement of direct solutions and approximated expressions with leading-order terms. Therefore, the higher-order correction terms that are neglected in deriving the Eqs. (\ref{Eq:19}) and (\ref{Eq:20}) are not visibly important here.     

\subsubsection{Effect of chiral inertial dynamics}
\begin{figure}[h]
    \centering
    \includegraphics[scale=0.42]{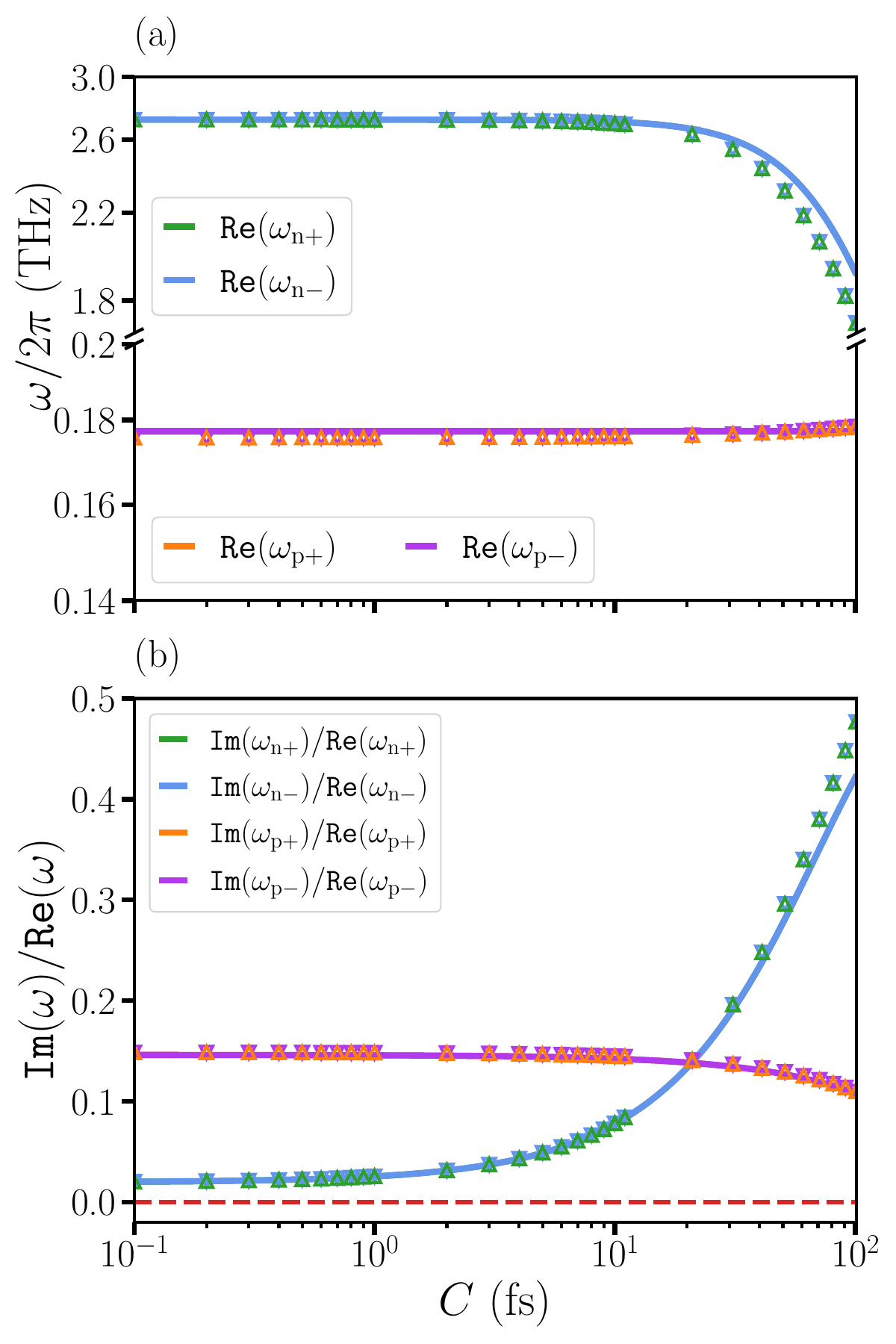}
    \caption{The variation of (a) precession and nutation frequencies and (b) effective damping plotted against $C$. The solid lines are for the analytical expressions derived in Eqs. (\ref{Eq:19}) and (\ref{Eq:20}), where as the symbols are the data { from the direct solutions of Eq. (\ref{Eq:12})}. The considered parameters are $M_{0}$ = $2\mu_B$, $\gamma$ = 28 GHz/T, $\alpha$= 0.05, $ J = 10^{-21}$ J, $K = 10^{-23}$ J, $D_z = 0$ J, $H_0 = 0$ T.}
    \label{fig:5}
\end{figure}
To understand the effect of chiral inertial dynamics, we fix $\mathbb{I}_{xx} = 0$, $\eta = 100$ fs, and vary the parameter $C$ in Eq. (\ref{Eq:12}). The direct solutions of Eq. (\ref{Eq:12}) have been presented in Fig. \ref{fig:5}. It can be observed that the nutation resonance frequencies decrease with $C$, while the precession resonance frequencies are almost constant with a very small increase at large $C$ values. Such an observation is consistent with the approximated expresseions in Eqs. (\ref{Eq:19}) and (\ref{Eq:20}). It can easily be seen that the dominant term in the expression of $\omega_{\rm p}$ in Eq. (\ref{Eq:19}) does not contain the chiral inertial parameter $C$. Therefore, it is expected that the precession resonance is unaffected by chiral inertial dynamics. The same can also be observed in the case of FM [see Fig. \ref{fig:3}]. The nutation resonance frequencies also remain constant upto the ratio of $C/\eta >$ 0.1. The reason is that the dominant term in $\omega_{\rm n}$ is $1/(\eta\sqrt{1 + (C/\eta)^2})$. The correction term here is $\sqrt{1 + (C/\eta)^2}$ that can provide a visible significance for the ratio $C/\eta>$ 0.1. Note that the direct solutions and approximated expressions differ at the regime $C/\eta>$ 0.1. This is in contrast to the case of anisotropic inertial dynamics in AFM. The reason is that we have considered only the leading order frequency terms in the perturbative expansion of $\omega_{(0)}$ in deriving the analytical expressions. The higher-order correction terms become very important in the analytical expressions derived in Eq. (\ref{Eq:20}). 
\begin{figure*}[tbh!]
    \centering
\includegraphics[scale = 0.4]{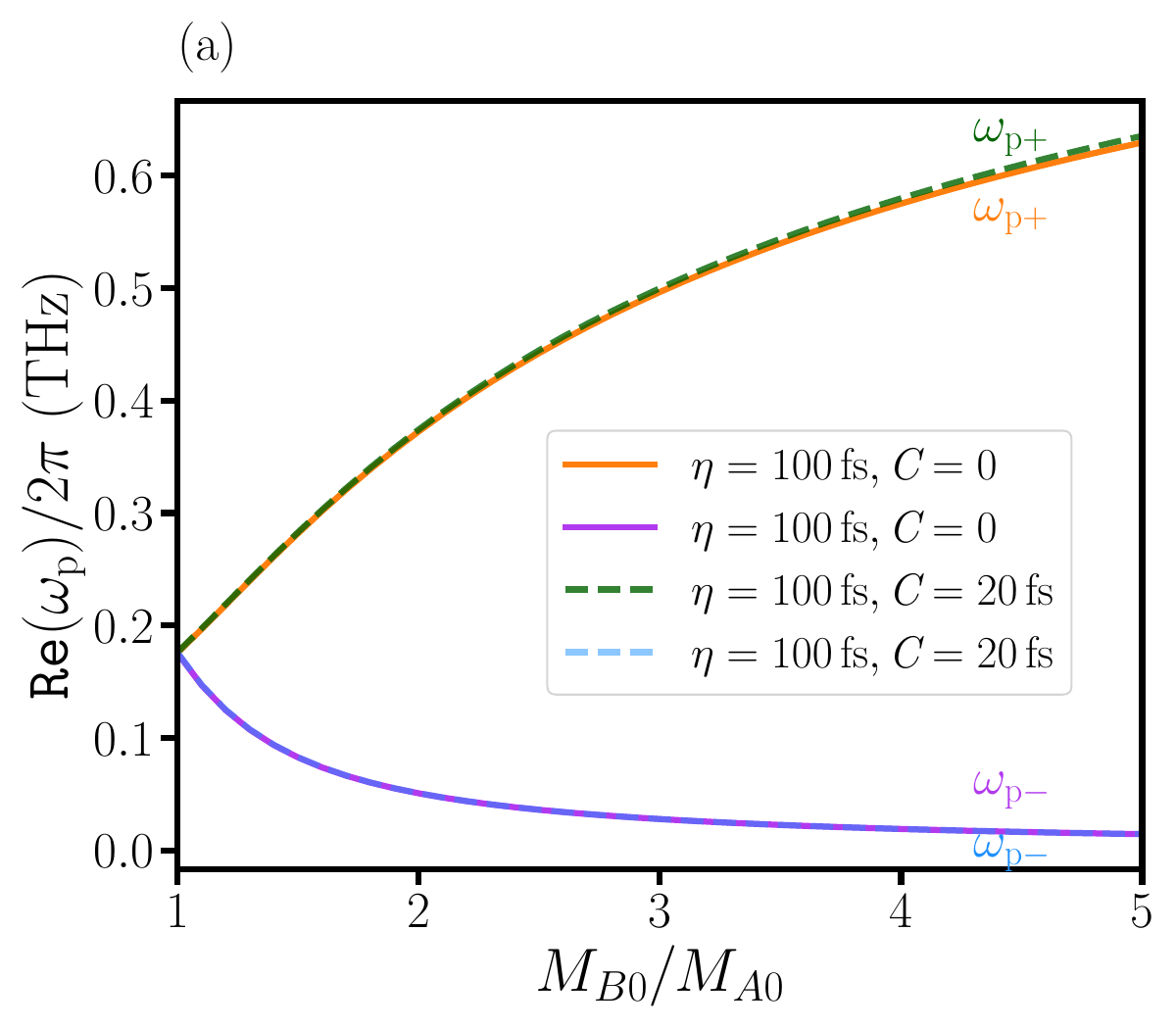}
\hspace{0.2cm}\includegraphics[scale = 0.4]{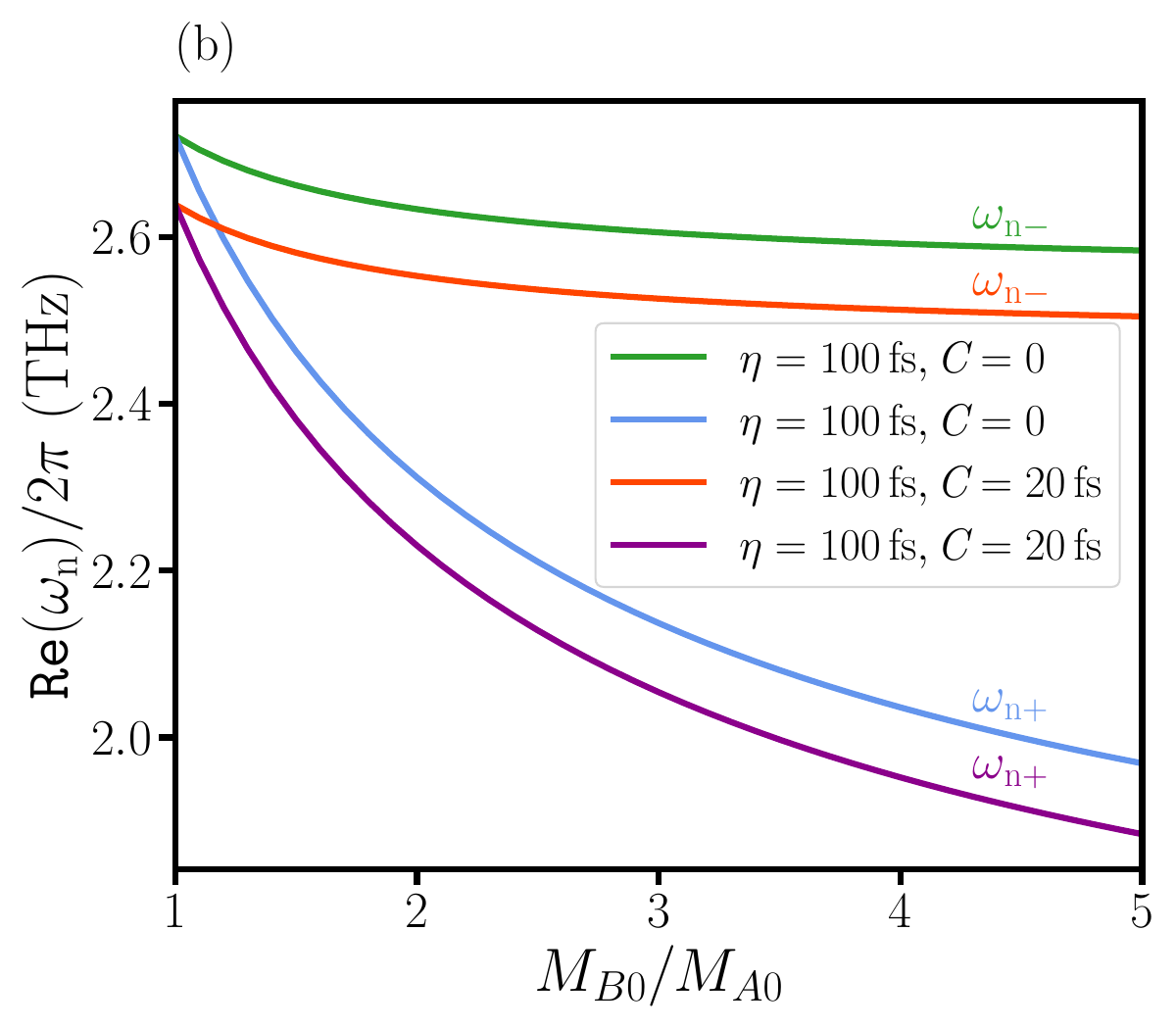}
    \caption{The variation of (a) precession and (b) nutation resonance frequencies as a function of the ratio of magnetic moments $M_{B0}/M_{A0}$ without and with the chiral inertial dynamics. The considered parameters are  $\gamma_A = \gamma_B$ = 28 GHz/T, $\alpha_A = \alpha_B$ = 0.05, $ J = 10^{-21}$ J, $K_A = K_B = 10^{-23}$ J, $D_z = 0$ J, $H_0 = 0$ T.}
    \label{fig:6}
\end{figure*}

While the chiral inertial time barely influences the precession resonance frequency, the effective damping of the precession resonance is reduced with increasing $C$. On the other hand, the effective damping of the nutation resonance increases quite sharply. This effect is consistent with the FM case as well. As the effective damping is increased, the kinetic energy introduced by the inertial dynamics will dissipate much faster, and hence the spin nutation effect will not last for a long time. Once again, we find disagreement of the computed effective damping between direct solutions and analytical solutions. The higher-order perturbation terms need to be incorporated into the analytical theory to agree with the direct solutions.   

\subsection{Ferrimagnets}\label{Section3C}

With the developed theory of tensorial inertial dynamics for AFM, we can further shed light on the FiM. The magnetic moments on the two oppositely aligned sublattices are different in FiM. The FiM already has two different precession frequencies. These are known as (i) ferromagnetic-like, (ii) antiferromagnetic-like or exchange mode \cite{Wienholdt2013,Wienholdt2012PRL,Marwan2022}. We compute these precession resonance frequencies, $\omega_{\rm p\pm}$ for FiM by solving Eq. (\ref{Eq:12}) in Fig. \ref{fig:6}(a). The ratio of magnetic moments $M_{B0}/M_{A0} = 1$ denotes for an AFM. As the ratio of magnetic moments increases, we observe that the resonance frequencies for the ferromagnetic-like precession mode (denoted by $\omega_{\rm p-}$) fall in the GHz regime. The resonance frequencies for the exchange mode (denoted by $\omega_{\rm p+}$)  escalate sharply and fall in the THz regime. It is visibly noticed that the chiral inertial time with $C = 20$ fs change the precession resonances within 0.93\%. The precession resonance frequencies calculated with and without the chiral inertial dynamics almost overlap.      

Due to the difference in magnetic moments, two nutation modes $\omega_{\rm n\mp}$ seperate  [Fig. \ref{fig:6}(b)], breaking the symmetry between $\omega_{\rm n+}$ and $\omega_{\rm n-}$. Since the precession and nutation resonances have opposite senses of rotation, the positive $\omega_{\rm p+}$ correspond to the negitive $\omega_{\rm n-}$ and vice versa.  \cite{Kikuchi}. For AFMs, we found that chiral inertial time decreases the nutation resonance frequencies; the same effect is observed in FiMs. Furthermore, a rapid decrease in $\omega_{\rm n+}$ can be seen, while the other nutation mode $\omega_{\rm n-}$ remains almost unaffected by the ratio of magnetic moments. To summarize, chiral inertial time does not significantly influence the precession resonances, but they decrease the frequencies of the nutation resonances in FiMs.

\section{Conclusions}\label{Section4}
The field of magnetic inertial dynamics has emerged as a possible way to manipulate ferromagnetic spins at THz frequencies. Several studies showed that magnetic inertia is a tensor \cite{Fahnle2011,Nagyfalusi2024,Thonig_2014}. Here we have formulated a linear response theory of ferromagnets and antiferromagnets with the magnetic inertia tensor. First, the inertia tensor has been decomposed into three parts: (i) scalar and isotropic inertia characterized by $\eta$, (ii) symmetric and anisotropic inertia tensor characterized by $\mathbb{I}$, and (iii) antisymmetric tensor and chiral inertia characterized by the vector $\bm{C}$. We find that the symmetric inertia tensor enters in the real part of the susceptibility, while the chiral inertia enters in the imaginary part of the susceptibility. Further, the effect of inertia tensor does not show prominent contributions in the precession resonance, however, the nutation resonance is affected significantly. Our results show overall reduction of nutation resonance frequencies with the increase of inertia tensor strength in FM and AFM. In contrast, the effective damping of the nutation resonance increases quite rapidly. The increase in nutation resonance damping affects the linewidth of the nutation resonance that might be detected in the experiments.

\section{Acknowledgments}
We thank Levente Rózsa, and Peter M. Oppeneer for fruitful discussions. Financial support by the faculty research scheme at IIT(ISM) Dhanbad, India under Project No. FRS(196)/2023-2024/PHYSICS and Startup Research Grant (SRG) by SERB Under Project No. SRG/2023/000612 are gratefully acknowledged.

\appendix
\section{Characteristic eigenmodes of precession and nutation resonances with tensorial inertia in ferromagnets}
Here, we compute the eigenmodes for the precession and nutation resonance in FM. The linear response theory applied to Eq. (\ref{eq:2}) results 
\begin{align}
   \begin{pmatrix}
h_x\\
h_y
\end{pmatrix} & =  \frac{1}{\gamma M_0} \begin{pmatrix}
\Omega_{0} +\alpha \partial_t + \eta^{\rm eff} \partial_{tt}  & - \partial _t + C \partial_{tt} \\
\partial _t -C \partial_{tt}  & \Omega_{0} +\alpha \partial_t + \eta^{\rm eff} \partial_{tt} 
\end{pmatrix}\nonumber\\& \cross
\begin{pmatrix}
m_x\\
m_y
\end{pmatrix}\,.
\label{A1}
\end{align}
Here we consider $\mathbb{I}_{xx} = \mathbb{I}_{yy}$ and $\mathbb{I}_{xy} = \mathbb{I}_{yx} = 0$. Note that we { write} the derivative{ s} in the following way: $\partial^2/\partial t^2 \equiv \partial_{tt}$ and $\partial/\partial t \equiv \partial_{t}$. Now we use $m_x, m_y \propto e^{{\rm i}\omega t}$ in Eq. (\ref{A1}) and obtain
\begin{align}
    \begin{pmatrix}
h_x\\
h_y
\end{pmatrix}  &=\frac{1}{\gamma M_0} \begin{pmatrix}
\Omega_{0} +{\rm i}\alpha\omega  - \eta^{\rm eff} \omega^2  & -{\rm i}\omega - C \omega^2 \\
{\rm i}\omega + C \omega^2  & \Omega_{0} +{\rm i}\alpha\omega  - \eta^{\rm eff}\omega^2 
\end{pmatrix}\nonumber\\& \cross
\begin{pmatrix}
m_x\\
m_y
\end{pmatrix}\,.
\label{A2}
\end{align}
To find the nature of eigenmodes we employ $h_x = h_y = 0$, $\alpha = 0$. Following Eq. (\ref{A2}), two coupled equations can be attained    
\begin{align}\label{A3}
    \left[\Omega_0- \eta^{\rm eff}\omega^2\right]m_x-\left({\rm i}\omega+C\omega^2\right)m_y=0\,,\\
    \left({\rm i}\omega+C\omega^2\right)m_x+\left[\Omega_0- \eta^{\rm eff}\omega^2\right]m_y=0\label{A4}\,.
\end{align}
To obtain the eigenmodes for precession resonance, we work with the leading-order term of $\omega_{\rm p}$ derived in Eq. (\ref{Eq7}) i.e. $\omega_{\rm p} \approx \Omega_0$. Eqs. (\ref{A3}) and (\ref{A4}) can be recast as   
\begin{align}\label{A5}
    m_x & \approx \left({\rm i}+C\Omega_0\right)m_y\,,\\
    m_y & \approx -\left({\rm i}+C\Omega_0\right)m_x \label{A6}\,.
\end{align}
\begin{figure}[h]
\centering
\includegraphics[scale=0.33]{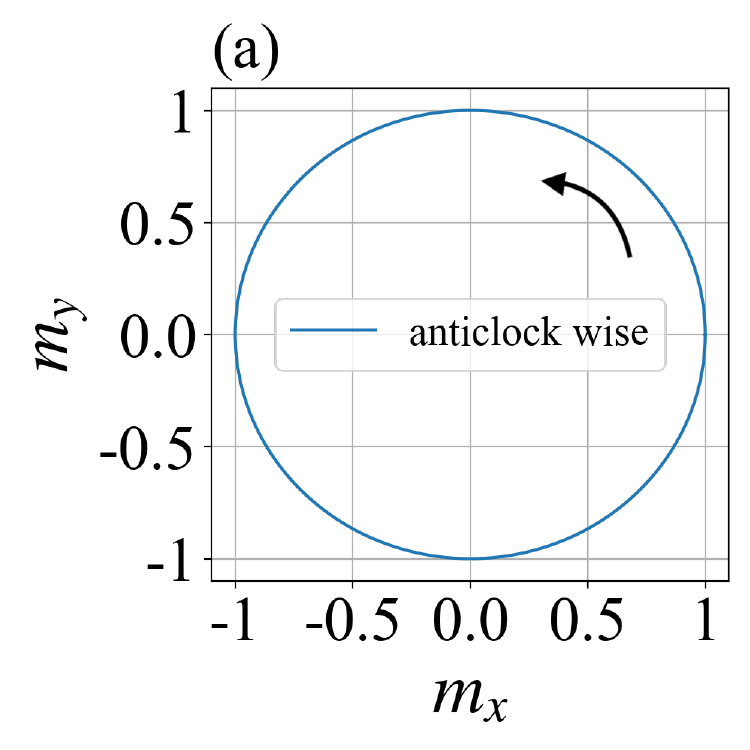}
\includegraphics[scale=0.33]{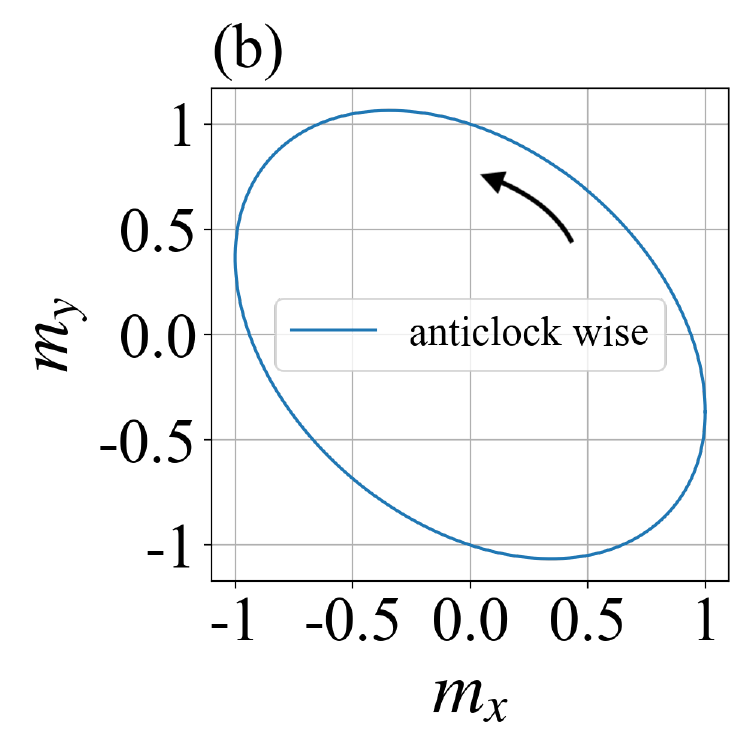}
\includegraphics[scale=0.33]{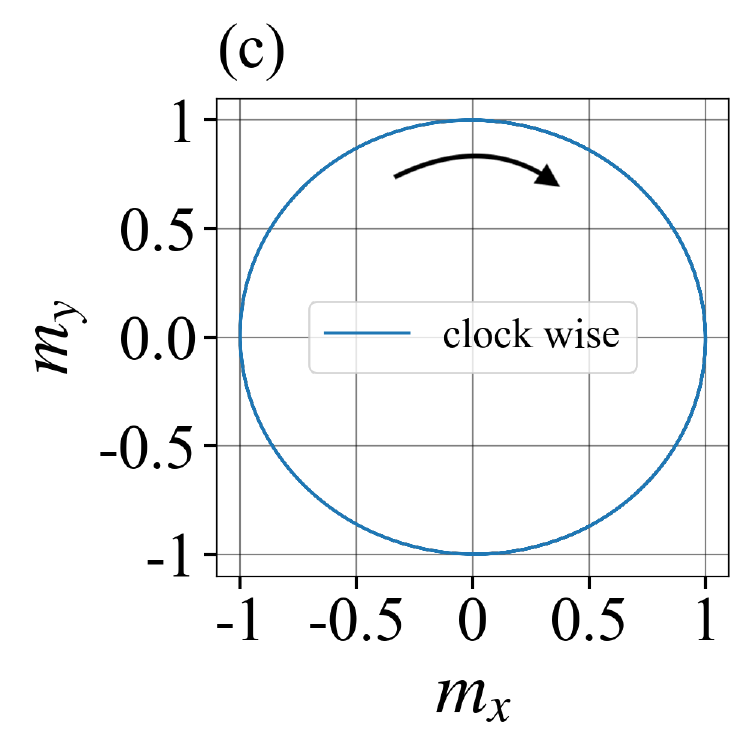}
\includegraphics[scale=0.33]{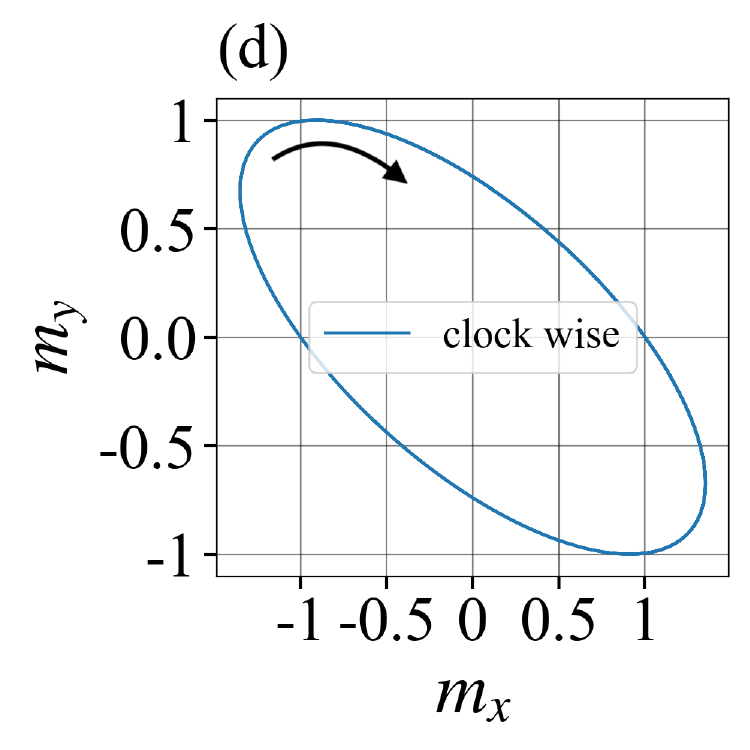}
\caption{Precessional resonance modes have been shown with (a) $C$ = 100 fs,  and (b) $C$ = 1 ps. Nutation resonance modes have been shown with (c) $C$ = 1 fs, $\eta$ = 100 fs, $\mathbb{I}_{xx}$ = 10 fs, and (d) with $C$ = 100 fs, $\eta$ = 100 fs, $\mathbb{I}_{xx}$ = 10 fs.}
\label{fig:S1}
\end{figure}
To obtain the time-dependent real parts of $m_x$ and $m_y$ we compute
\begin{align}\label{A7}
    m_x & = {\tt Re}\left[m_xe^{{\rm i}\omega t}\right] = {\tt Re}\left[m_xe^{{\rm i}\Omega_0 t}\right] = m_x \cos\Omega_0 t\,,\\
    m_y & = {\tt Re}\left[m_ye^{{\rm i}\omega t}\right] = {\tt Re}\left[-m_x\left({\rm i}+C\Omega_0\right)e^{{\rm i}\Omega_0 t}\right]\nonumber\\&
    =m_x\sin\Omega_0 t - C\Omega_0m_x\cos\Omega_0 t
    \label{A8}\,.
\end{align}
We can see that the precession resonance eigenmode depends on the chiral magnetic inertia $C$. Such eigenmodes are shown for two values of $C$ in Figs. \ref{fig:S1}(a) and \ref{fig:S1}(b). We find that the eigenmodes become 
 elliptic due to the higher value of chiral inertial time $C$.

Now we turn to the eigenmodes of nutation resonance. We substitute the leading order term in $\omega_{\rm n} \approx-\frac{\eta^{\rm eff}}{\left(\eta^{\rm eff}\right)^2 + C^2} $ and obtain the relation between $m_x$ and $m_y$ as,
\begin{align}\label{A9}
    m_x  
    &
    =\frac{-{\rm i}\left(\eta^{\rm eff}\right)^3 - {\rm i} \eta^{\rm eff} C^2 +C\left(\eta^{\rm eff}\right)^2}
{\Omega_0\left[\left(\eta^{\rm eff}\right)^2 + C^2\right]^2 - \left(\eta^{\rm eff}\right)^3}m_y \nonumber\\
& = \frac{-{\rm i} \left(1+ \left[\frac{C}{\eta^{\rm eff}}\right]^2\right) + \frac{C}{\eta^{\rm eff}}}{\Omega_0 \left(\eta^{\rm eff} + 2C \frac{C}{\eta^{\rm eff}} + C \left[\frac{C}{\eta^{\rm eff}}\right]^3 \right) - 1} m_y\,.
\end{align}
\begin{figure*}[tbh!]
\centering
\includegraphics[scale=0.40]{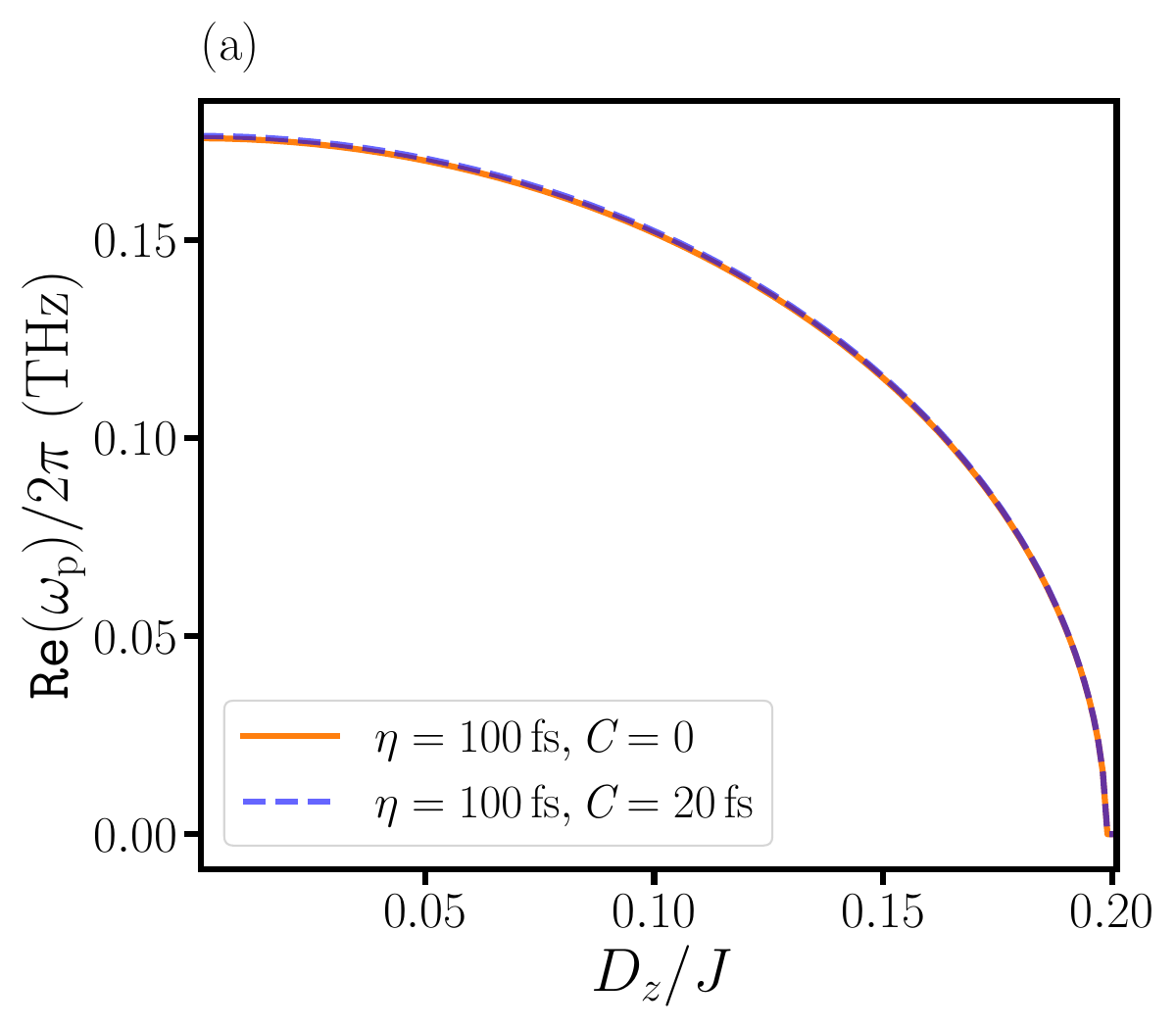}
\includegraphics[scale=0.40]{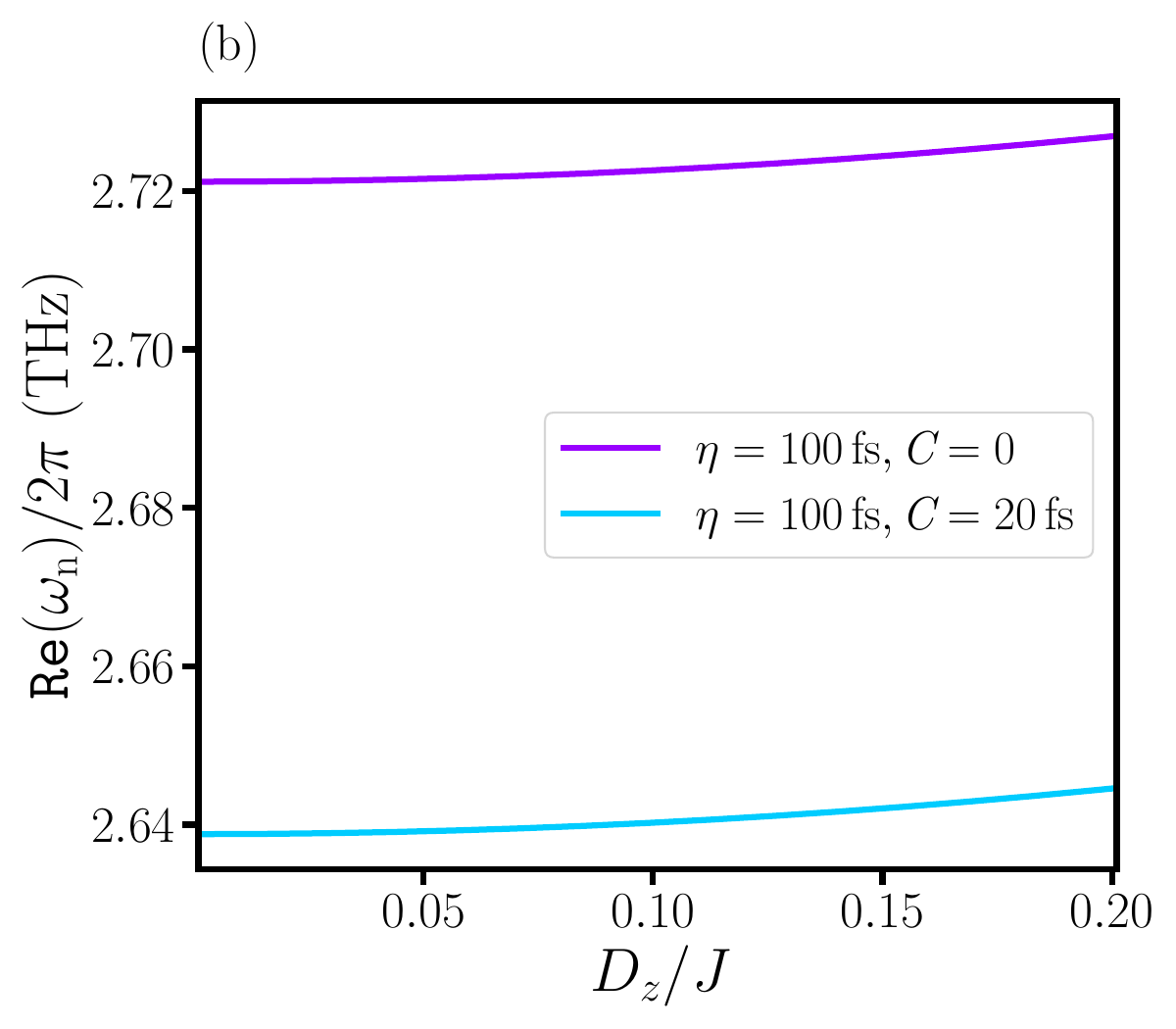}
\caption{Variation of (a) precession and (b) nutation resonance frequencies as a function of the ratio of DMI strength and the exchange interaction strength i.e., $D_z/J$  with and without chiral inertia for AFM. The considered parameters are  $\gamma_A = \gamma_B$ = 28 GHz/T, $\alpha_A = \alpha_B$ = 0.05, $ J = 10^{-21}$ J, $K_A = K_B = 10^{-23}$ J, $H_0 = 0$ T.}
\label{fig:S2}
\end{figure*}
In the limit $C/\eta^{\rm eff} \ll 1$, we find the following relations 
\begin{align}
    m_x \approx  {\rm i}m_y\,,\\
    m_y\approx-im_x\,.
\end{align}
The real and time-dependent part of $m_x$ and $m_y$ at nutation resonance is calculated as
\begin{align}\label{34}
    m_x & = {\tt Re}\left[m_xe^{{\rm i}\omega t}\right]= {\tt Re} \left[m_xe^{-{\rm i}\left(\frac{\eta^{\rm eff}}{\left(\eta^{\rm eff}\right)^2 + C^2} \right)t }\right]\nonumber\\&
    = m_x 
\cos\left[\frac{\eta^{\rm eff}}{\left(\eta^{\rm eff}\right)^2+C^2}\right] t\,,\\
    m_y & = {\tt Re}\left[m_ye^{i\omega t}\right] = {\tt Re}\left[-im_xe^{-{\rm i}\left(\frac{\eta^{\rm eff}}{\left(\eta^{\rm eff}\right)^2 + C^2} \right)t }\right]\nonumber\\&= - m_x \sin\left[\frac{\eta^{\rm eff}}{\left(\eta^{\rm eff}\right)^2+C^2}\right] t\,.
\end{align}
These characteristic nutation resonance modes have been shown in Fig \ref{fig:S1}(c) for the limit $C/\eta^{\rm eff} \ll 1$. 

However, the eigen mode of nutation resonance change when $C/\eta^{\rm eff} \approx 1$. To realize this, we employ $\Omega_0 \eta^{\rm eff}\ll 1$ and Eq. (\ref{A9}) takes the form
\begin{align}
    m_x & \approx  \left[{\rm i} \left(1+ \left[\frac{C}{\eta^{\rm eff}}\right]^2\right) - \frac{C}{\eta^{\rm eff}}\right] m_y\,.
\end{align}
Using the above relation we compute the nutation resonance eigen mode for $C = 100$ fs, $\eta^{\rm eff} = 110$ fs. The eigen mode has been shown in Fig. \ref{fig:S1}(d).

\section{Effect of DM interaction on chiral inertial dynamics}

To understand the effect of DMI on the tensorial inertial dynamics, we compute the precession and nutation resonance frequencies as a function of the ratio of DMI strength $D_z$ and Heisenberg exchange interaction strength $J$ i.e., $D_z/J$. We directly solve Eq. (\ref{Eq:12}) and calculate $\omega_{\rm p}$ and $\omega_{\rm n}$. It is clear from Eq. (\ref{Eq:19}) that a realistic precession frequency demands $4K(J+K)>D_z $. Therefore, we restrict to such conditions. The results are plotted in Figs. \ref{fig:S2}(a) and  \ref{fig:S2}(b). The precession resonance frequencies reduce with the DMI, however, the chiral inertial time do not change the precession resonance. On the other hand, the nutation resonance frequencies remain almost constant even though $D_z$ is introduced. The chiral inertial time decreases the nutation resonance frequencies.      


\providecommand{\noopsort}[1]{}\providecommand{\singleletter}[1]{#1}%
\end{document}